# A Hybrid CNN-LSTM Approach for Laser Remaining Useful Life Prediction

Khouloud Abdelli[1,2], Helmut Grießer[1], and Stephan Pachnicke[2]

[1]*ADVA Optical Networking SE, Fraunhoferstr. 9a, 82152 Munich/Martinsried, Germany*
[2] *Christian-Albrechts-Universität zu Kiel, Kaiserstr. 2, 24143 Kiel, Germany*
*E-mail: KAbdelli@adva.com*

**Abstract:** A hybrid prognostic model based on convolutional neural networks (CNN) and long short-term memory (LSTM) is proposed to predict the laser remaining useful life (RUL). The experimental results show that it outperforms the conventional methods. © 2021 The Authors

## 1. Introduction

Remaining useful life (RUL) is generally defined as the time remaining for a component to perform its intended function before failing. Accurate laser RUL estimation is crucial for condition-based maintenance to enhance reliability, to prevent unplanned downtime and thereby to minimize the maintenance costs.

RUL estimation approaches can be broadly categorized into model-based (physics-based) and data-driven approaches. Physics-based techniques use explicit mathematical equations to model the degradation behavior of the system by incorporating domain knowledge and the understanding of the system's physical phenomena. Although they are precise and do not require plenty of data, they are time-consuming, potentially difficult to implement and hard to apply to complex systems. Data-driven approaches, extracting useful insights from the operational data collected to learn the degradation trend and thus to perform the prognostics without requiring any specific knowledge or using any physical model, have gained popularity with the rapid advances in data collection, storage, and mining techniques. In this respect, Fan et al. [1] presented a particle filter-based prognostic approach based on both, sequential Monte Carlo and Bayesian techniques to predict the lumen maintenance life of LED light sources. Liu et al. [2] proposed a data-driven similarity-based difference analysis approach for RUL prediction of GaAs lasers. We investigated a data-driven fault detection model based on Long Short-Term Memory (LSTM) to detect the laser degradation [3].

In this paper, we propose a novel data-driven prognostic approach integrating convolutional neural networks (CNN) and LSTM to estimate the RUL of vertical-cavity surface-emitting laser (VCSEL) by taking the output power time series extended with different laser parameters as the inputs. The integration of CNN and LSTM by combining their strengths would help to significantly extract the hidden patterns in the laser degradation data with multiple operating conditions and thus improve the performance of the model. The proposed approach, called CNN+LSTM in the following, is applied to real laser reliability data derived from accelerated aging tests carried out at different conditions. The experimental results show that our model (i) achieves more accurate RUL estimation compared to several baseline ML techniques, namely support vector regression (SVR), random forest (RF), multi-layer perceptron (MLP), CNN, and LSTM and (ii) outperforms the conventional laser RUL estimation methods based on linear least squares fitting.

## 2. Setup & Configurations

### 2.1. Accelerated Aging Tests

Different accelerated aging tests for VCSEL samples with various oxide aperture sizes under multiple controlled operating conditions, namely the current $I$, the junction temperature $T_j$, the current density $J$ and the resistance $R$, are performed up to 3,500 hours. The temperature $T$ is varied from 85°C to 150°C to strongly increase the laser degradation and thus accelerating the device failure. The current is held constant for the duration of the aging test. The optical output power is monitored periodically. The time to failure $t_f$ of the device is defined as the time at which the output power has decreased by 1 dB (20%) of its initial value. It is calculated using a linear interpolation between the last output power measurement above -1 dB and the first output power value below -1 dB. Figure 1 shows an example of life test results of VCSEL devices conducted at 125°C. In Fig. 1 the VCSEL devices are tested under constant current of 11 mA, and the operating output power is used as the degradation parameter for the $t_f$ estimation. As depicted in Fig. 1, the output power shows a significant decrease with increasing aging test time.

### 2.2 Data Set Preparation

The output power measurements of the different devices are recorded from the beginning of the aging test until $t_f$ of the device or the end of the test if the device has not failed during the test. The power measurements of the devices, which failed during the first 100 hours of aging tests due to manufacturing defects, are not taken into consideration for the ML model training. In total, 239 output power time series are generated. To increase the amount of data for the training of the ML model, the time series are split with a sliding window of size 3. The derived sequences are combined with the operating conditions, namely the oxide aperture size (OA), $T_j$, $I$, $T$, $J$ and $R$. The RUL of each sequence is estimated as the difference between $t_f$ and the time $t$ at which the RUL is predicted. As it is impossible to predict the RUL for very long time periods, only the measurement sequences recorded before the failure happened up to a maximum of 5,000 hours are considered for the training of the ML model. In total, a data set of 729 samples is built. The generated data is normalized and divided into a training (comprising of 80% of the sequences) and a test dataset (the remaining 20%).

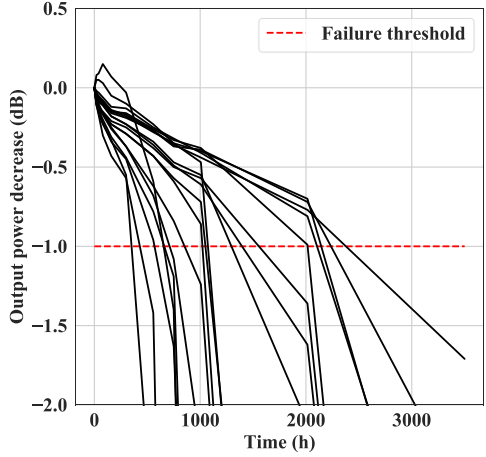
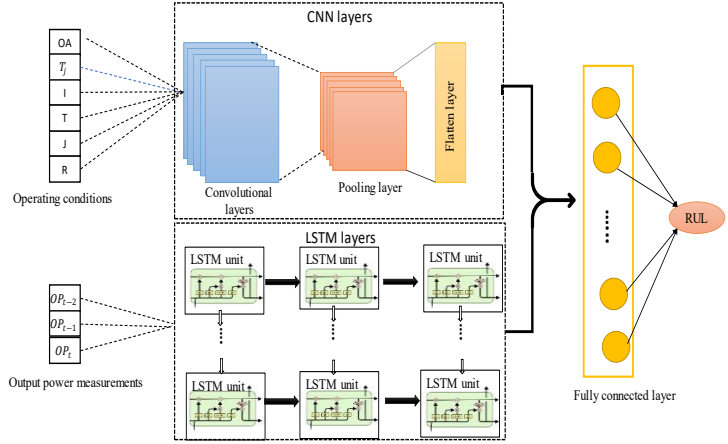

Fig. 1: Experimental aging data of VCSELs conducted at 125°C    Fig. 2: Proposed model architecture.

### 2.2 Machine Learning Model

As shown in Fig. 2, the proposed framework consists of two parallel models (one based on CNN and one on LSTM) followed by a fully connected neural network combining the outputs of each model to achieve the RUL estimation regression task. As a CNN has a good ability to extract local spatial features, the operating conditions are fed to the CNN based model composed of the three stacked convolution layers containing 32, 32 and 16 filters, respectively, followed by max pooling layers. Since LSTM is appropriate to capture the dependency within sequential data, the output power measurement sequence is provided to the LSTM based model composed of three stacked LSTM layers made up of 40, 20 and 10 memory cells, respectively. The 2-dimensional output of the CNN model is flattened into a one-dimensional vector before combining it with the output of the LSTM model. The fused output is then used as an input to the neural network layer consisting of 64 neurons which outputs the predicted RUL.

The accuracy of the ML model RUL estimation is quantified by using several evaluation metrics, the root mean square error (RMSE), the mean absolute error (MAE), and the scoring function S [4] which is defined as,

$$S = \begin{cases} \sum_{i=1}^{N} \left( e^{-\frac{RUL_{pred(i)} - RUL_i}{a_1}} - 1 \right), & if\ RUL_{pred(i)} < RUL_i \\ \sum_{i=1}^{N} \left( e^{\frac{RUL_{pred(i)} - RUL_i}{a_2}} - 1 \right), & if\ RUL_{pred(i)} \geq RUL_i \end{cases} \quad (1)$$

where $N$ is the total number of samples in the test data set and $RUL_i$ and $RUL_{pred(i)}$ denote the true RUL and the predicted RUL for the test sample $i$, respectively. The parameters $a_1$ and $a_2$ are user-defined parameters controlling the asymmetric preference of underestimated predictions over overestimated predictions. For our experiments, $a_1$ and $a_2$ are set to 250 and 220.

The RMSE and MAE metrics are used to measure how close the predicted RULs are to the true RULs by equally penalizing the underestimated and overestimated predictions. Whereas the scoring metric tends to highly penalize the cases when the predicted RUL is larger than the actual RUL since overestimated values may lead to device failure resulting in higher costs.

### 3. Results and Discussions

We compare the performance of our proposed model CNN+LSTM with several other ML algorithms, like SVR, RF, MLP, CNN and LSTM. As shown in Fig. 3, our model significantly outperforms the aforementioned ML techniques in terms of all metrics by achieving the smallest values of RMSE, MAE and scoring. It achieves a notable improvement over the different ML approaches particularly SVR and RF. Specifically, the improvement on scoring is more than

26.8% whereas the improvements on the RMSE and MAE metrics are more than 11.5% and 16% respectively. To examine the RUL estimation capability of our model, we compare the predicted RUL values to the true RULs at different stages of degradation. As shown in Fig.4, the estimated RUL values are closer to the actual RULs (the standard deviation $\sigma = 0.3$) when the devices approach end of life (i.e RUL $\leq 100\ h$), and farther away ($\sigma = 0.5$) during the early-stage degradation phase (i.e RUL $\geq 2000\ h$). The gap between the predicted and actual RUL values can be mainly explained by the limited training data that affected the performance of the ML Model.

To showcase the RUL prediction performance along the lifespan, the true and predicted RUL values of a single selected laser device are compared. Figure. 5 shows that the prediction tends to be very close to the true RUL when the device is approaching end of the life.

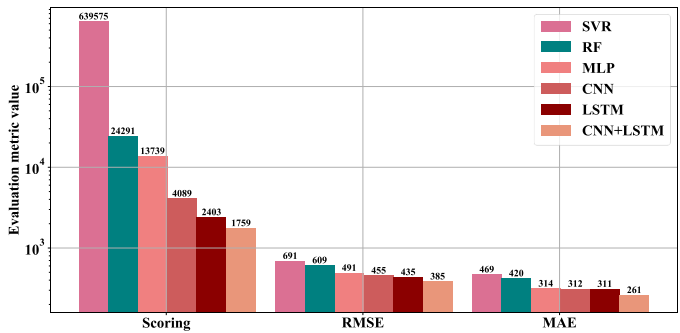 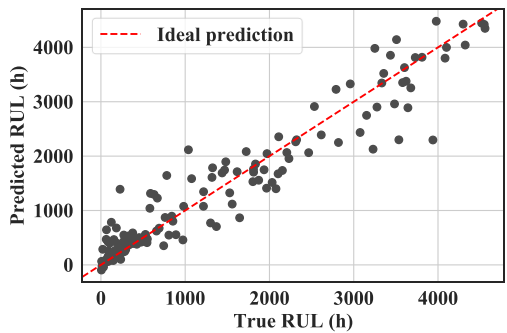

Fig. 3: Comparison of the results of our model with other methods    Fig. 4: Predicted RULs by our model vs. actual RULs

The performance of the proposed model is compared with a conventional laser RUL estimation technique based on a linear least-squares-fitting (LLSF) method in terms of prediction error. The comparison of the results shown in Fig. 6 demonstrates that our framework outperforms the conventional method significantly by achieving smaller prediction errors. The LLSF technique significantly underestimates or overestimates the RUL whereas the estimates of CNN+LSTM are closer to the true RULs.

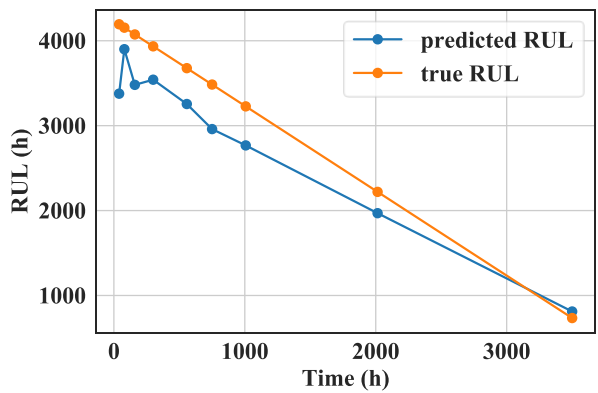 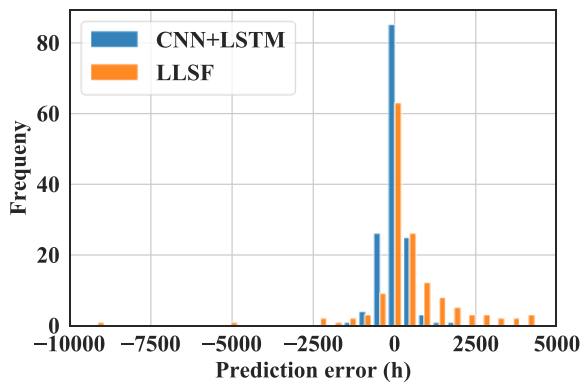

Fig. 5: The true and predicted RULs of an example laser device    Fig. 6: Histogram of prediction errors of CNN-LSTM and LLSF

## 4. Conclusion

We proposed a new hybrid deep learning approach for laser RUL estimation. The presented framework is based on the integration of a deep CNN and a deep LSTM model coupled via fusion and a fully connected neural network to enhance the performance. The proposed model has been tested using real laser reliability data modelling 2.1the laser degradation under different operating conditions. The results demonstrated that the proposed method outperforms different ML algorithms as well as the conventional technique. Future work will focus on increasing the performance of our model by collecting more data.

The work has been partially funded by the German Ministry of Education and Research in the project OptiCON (#16KIS0989K).